\documentstyle[aps,multicol,epsf,epsfig,aps,rotate]{revtex}

\begin{document}

\draft

\title{The Continuum Percolation Threshold \\ for Interpenetrating
Squares and Cubes}

\author{Don R. Baker,$^{\ast\dagger}$ Gerald Paul,$^\ast$ Sameet
 Sreenivasan,$^\ast$ and H. Eugene Stanley$^\ast$}

\address{$^\ast$Center for Polymer Studies and Dept. of Physics,
Boston University, Boston, MA 02215 USA\\ 
$\dagger$Department of Earth and Planetary Sciences,
McGill University\\ 3450 rue University, Montr\'eal, QC H3A 2A7 Canada}

\date{11 March 2002}

\maketitle

\begin{abstract}

Monte Carlo simulations are performed to determine the critical
percolation threshold for interpenetrating square objects in two
dimensions and cubic objects in three dimensions.  Simulations are
performed for two cases: (i) objects whose edges are aligned parallel to one another and (ii) randomly oriented objects.  For squares whose
edges are aligned, the critical area fraction at the percolation
threshold $\phi_c=0.6666\pm 0.0004$, while for randomly oriented squares
$\phi_c=0.6254\pm 0.0002$, 6\% smaller. For cubes whose edges are
aligned, the critical volume fraction at the percolation threshold
$\phi_c=0.2773\pm 0.0002$, while for randomly oriented cubes
$\phi_c=0.2236\pm 0.0002$, 24\% smaller.

\end{abstract}

\begin{multicols}{2}

\section {Introduction}

Lattice percolation is often used for the statistical modeling of
transport in porous media
\cite{Ben-Avraham00,Bunde96,Sahimi94,SA94}. The requirement that sites,
and the bonds between them, be constrained to a fixed lattice may not,
however, be an appropriate model for natural porous media
\cite{Ben-Avraham00,Bunde96,Sahimi94,SA94}.  The characteristics of site
and bond percolation potentially limit their applicability to modeling
of natural phenomena such as oil or groundwater flow and extraction of
melt from super-solidus regions deep inside a planetary body.

Continuum percolation offers two advantages for describing porous media:

\begin{itemize}

\item[{(i)}] The objects which form clusters are not restricted to
points on a fixed lattice; they can be placed anywhere within the volume
studied and either be barred from interpenetration, or allowed to
interpenetrate, i.e., they can have either ``hard'' or ``soft'' cores
\cite{SA94}.  Because of the freedom of placement inside the system, the
connections between soft core objects can range from very small to very
large, depending upon the extent of interpenetration.

\item[{(ii)}] The objects can be of any shape. In two dimensions the
continuum percolation of discs is often investigated
\cite{Bunde96,SA94}.  In most studies of continuum percolation in three
dimensions, spheres are used as the objects, leading to the
``swiss-cheese'' nomenclature for continuum percolation
\cite{Bunde96,SA94}. Other frequently-used shapes are rods and
ellipsoids of revolution \cite{Garboczietal95}.  In a few cases the
continuum percolation of cubes has been considered \cite{Alonetal90}.

\end{itemize}

Here we determine the threshold for continuum percolation of soft core
squares in two dimensions (2-D) and cubes in three dimensions (3-D)
whose edges are aligned parallel to, or oriented at random angles to,
the axes of the system.  Continuum percolation is believed to belong to
the same universality class as site and bond percolation
\cite{GawlinskiStanley81,GeigerStanley82}; once we have determined the
continuum threshold for an object of a specific shape we can apply many
of the characteristics of site and bond percolation, e.g. critical
exponents, to describe the continuum percolation cluster.

\section{Methods}

We construct 2-D and 3-D Monte Carlo simulations for the
determination of the percolation threshold based upon the Leath method
\cite{Leath76} and the methods Lorenz and Ziff \cite{Lorenz01} used
in their study of the continuum percolation of spheres. 

\begin{figure}
\centerline{
\epsfxsize=8.0cm
\epsfclipon
\epsfbox{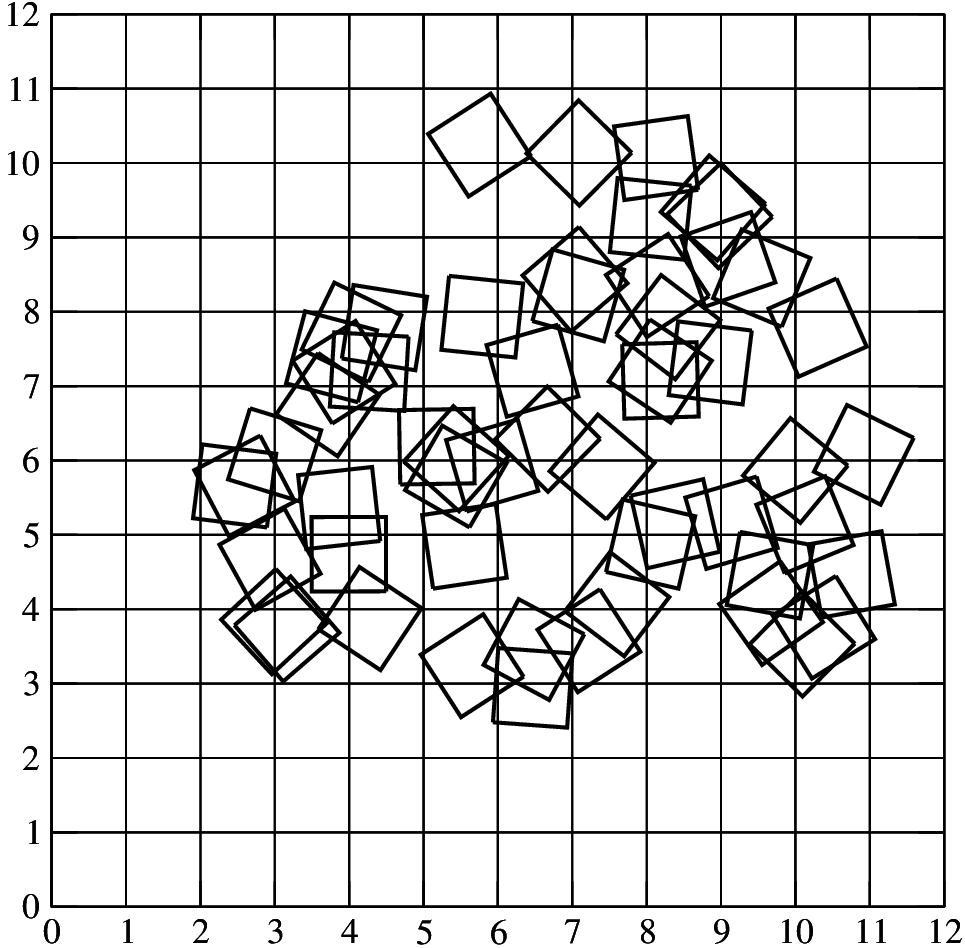}
}
\caption{Two-dimensional, $12\times 12$ example of a percolation cluster
of unit-length square objects (thick lines) for the case of randomly
oriented squares.  The system is divided by a
series of grid lines (thin lines) that create unit areas in this 2-D
system.  Note that the upper two objects of the cluster in the center of
the system intersect each other even though their centers are placed in
next-nearest-neighbor areas of the grid. The real 2-D and 3-D systems of
our study are much larger than this system.}
\label{Figure1}
\end{figure}

We perform 2-D simulations with squares whose edges are of unit length.
In 3-D we perform simulations with cubes of three different edge
lengths: $1/\sqrt 3$, 0.75, and 1.0; the length $L$ of each axis of the
simulation box in the 2-D simulations is 301, and in 3-D is 101 in
simulations using cubes with edge lengths of $1/\sqrt 3$ and 0.75, and
is 161 in simulations with the unit cube \cite{Text1}. We subdivide the
system into either a 2-D or 3-D grid of unit area squares or unit volume
cubes; an illustrative $12\times 12$ 2-D version of our system is shown
in Fig.~\ref{Figure1}.

The cluster begins in the center grid volume and objects are added to it
based upon a Poisson distribution centered about the average number of
objects per unit area or volume, $N/L^d$, chosen for the simulation of
dimension $d$ \cite{Lorenz01}.  The product of this value and the
individual object's area or volume, $v$, is the reduced number density
\begin{equation}
\eta\equiv\frac{vN}{L^d}.
\label{equation1}
\end{equation}
If the number of objects, $n$, generated from the Poisson distribution
is nonzero, then these $n$ objects are placed at random locations inside
the grid volume. To fix the orientation of each individual object,
random numbers are generated to determine the one angle of rotation in
the 2-D simulations and the three Euler angles of rotation in the 3-D
simulations; these angles vary between 0 and 2$\pi$.  The locations of
the 4 (2-D) or 8 (3-D) corners and the center of each object are stored
in a data structure, along with a flag indicating that the grid area
(2-D) or volume (3-D) was visited and populated during the
realization. The nearest-neighbors and next-nearest-neighbors of this
grid area or volume are then populated in a similar manner and the
intersection between squares or cubes is tested.

To determine whether two squares intersect we choose one square in the
cluster as the reference square and another square in the simulation as
the test square. We use an algorithm for the intersection of two
lines\cite{bourke89} to test if any of the 4 edges of the reference
square interesect the 4 edges of the test square.  To determine if two
cubes intersect we choose one cube in the cluster as the reference cube
and another in the simulation as the test cube.  We use an algorithm for
the intersection of a line and a facet \cite{bourke97} to test if any of
the edges of the reference cube intersect the faces of the test cube.
In this algorithm, the location of each of the 12 edges of the reference
cube are compared to the location of the 12 triangular facets that
describe the locations of all faces on the test cube using the corners
and diagonal of each face. If the test object intersects the reference
object, it is added to the growing cluster.  This process is repeated
for each new square or cube added to the system until the cluster can no
longer grow.  Intersections between squares or cubes in grid areas or
volumes up to two units away can occur for edge lengths 0.75 and 1, as
exemplified in 2-D for squares of unit edge (Fig.~\ref{Figure1}), but
cubes of edge $1/\sqrt 3$ can only interesect if they are in the same or
neighboring volumes of the grid, which reduces the number of grid
volumes that must be checked for cube overlap in simulations with cubes
of this smallest size.

The cumulative distribution of cluster sizes is calculated from the
cluster size of each realization, $s$, by binning the cluster sizes such
that all bins in the range of $2^0$ to $2^{s+1}-1$ are incremented by 1.
In order to estimate finite-size effects of the simulation, objects in
each cluster are tested to determine whether they touch the edge of the simulation. If so,
the cluster size is compared against the smallest cluster size in
previous realizations that touch the edge and the smaller value is
stored. Bins of size greater than the smallest cluster that touched the
edge of the system are not used in the determination of the percolation
threshold.

At the end of the simulation, the value of each bin is divided by the
number of realizations, from 10,000 to 50,000, to yield the probability
of achieving a cluster of size $s$, $P(s|\eta)$, for a given value of
$\eta$.  Power law behavior of the probability as a function of the bin
size is interpreted to indicate the critical percolation theshold
$\eta=\eta_c$ \cite{Bunde96,SA94}.  To accurately determine the
theshold, we follow the techniques of Ref.~\cite{Lorenz01}.  The
probability of generating a cluster size $s$ at a specified $\eta$ is
\cite{Bunde96,SA94}
\begin{equation}
P(s|\eta) \sim  As^{2-\tau}f[(\eta-\eta_c)s^{\sigma}],
\label{equation2}
\end{equation}
where both $\tau$ and $\sigma$ are universal exponents and $A$ is a
non-universal constant.  In 2-D the values of these exponents are 187/91
and 36/91 \cite{SA94}, respectively.  In 3-D the values of $\tau$ and
$\sigma$ are $2.18906\pm 0.00006$ and $0.4522\pm 0.0008$, respectively
\cite{Ballesteros99}.  Near the percolation threshold the scaling
function $f(x)$ can be expanded in a Taylor series
\begin{equation}
f(x)= 1 + Bx+{\cal O}(x^2).
\label{equation3}
\end{equation}
Combining Eqs.~(\ref{equation2}) and (\ref{equation3}),
\begin{equation}
P(s|\eta)s^{\tau-2}\sim  A + AB(\eta - \eta_{c})s^{\sigma}+\cdots
\label{equation4}
\end{equation}
which demonstrates that $ P(s|\eta)s^{\tau-2}$ becomes constant at the
percolation threshold as $s$ becomes asymptotically large.

\section{Results}

The percolation threshold can be expressed either as the critical
reduced number density, $\eta_c$, or the critical area (or volume)
fraction, $\phi_c$, which are related to each other by
\cite{ShanteKirkpatrick71}
\begin{equation}
\phi_c=1-e^{-\eta_c}.
\label{equation5}
\end{equation}

\subsection{Two Dimensions}

\begin{figure}
\centerline{
\epsfxsize=8.0cm
\epsfclipon
\epsfbox{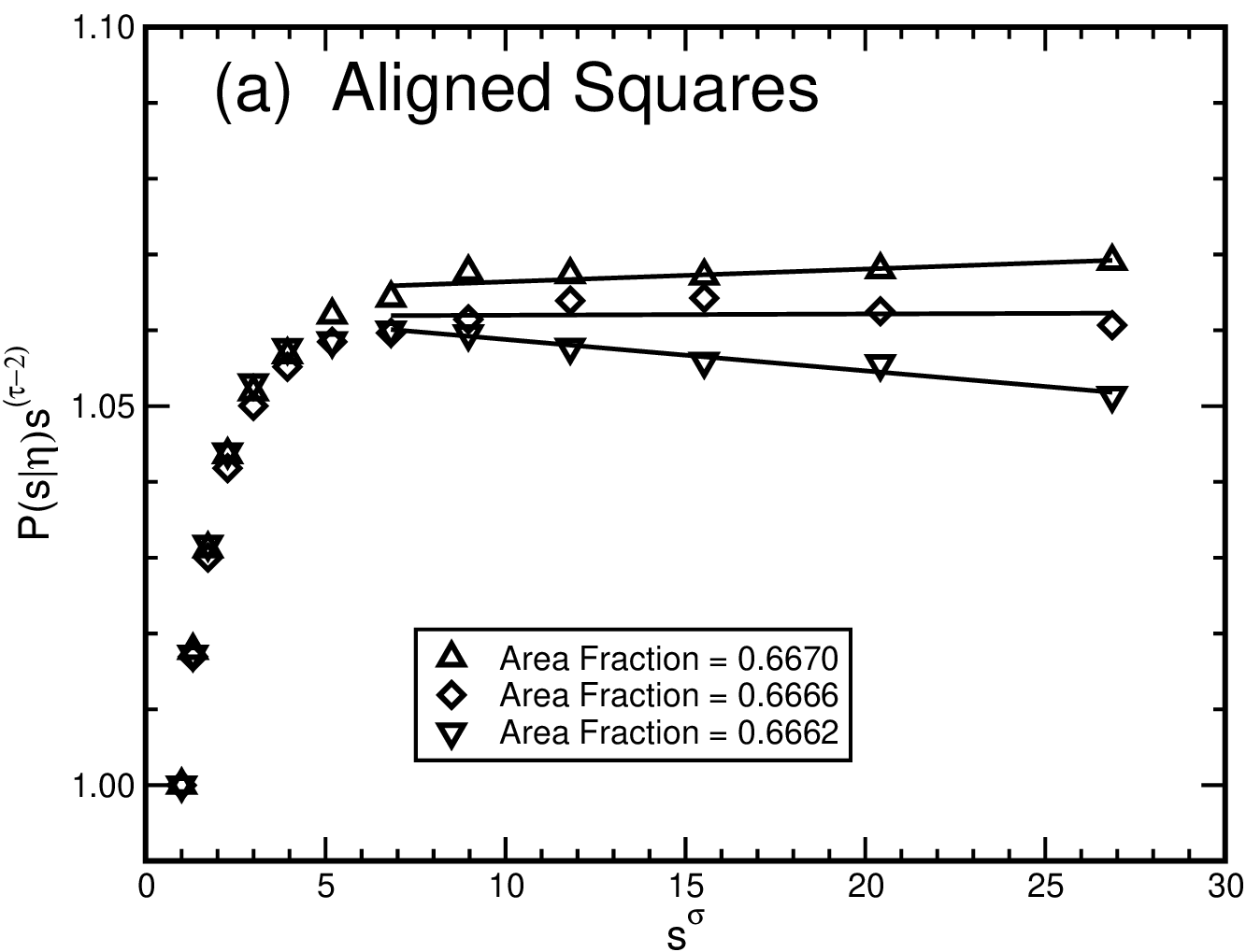}
}
\centerline{
\epsfxsize=8.0cm
\epsfclipon
\epsfbox{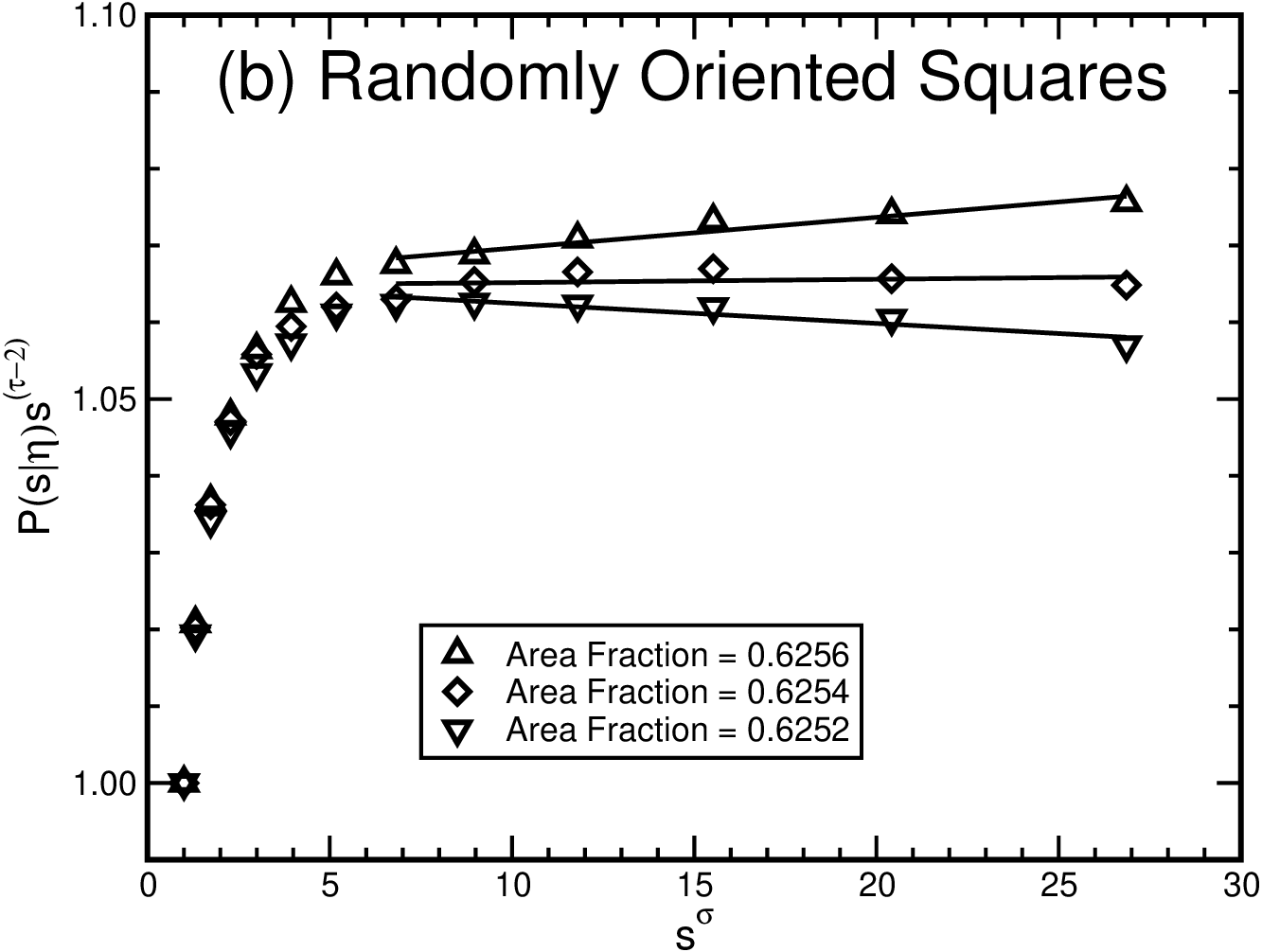}
}
\caption{Power-law scaled plots for determination of percolation
threshold for squares of unit size in a $301\times 301$ system based
upon 50,000 realizations at each area fraction.  At the threshold
$\eta=\eta_c$, $P(s|\eta)s^{\tau-2}$ is independent of $s^\sigma$, which
allows for accurate determination of $\eta_c$, which is related to
$\phi_c$ by Eq.~\protect\ref{equation5}. (a) Squares whose edges are
aligned parallel to each other, for which case we estimate
$\eta_c=1.098\pm 0.001$, so $\phi_c=0.6666\pm 0.0004$ by
Eq.~\protect\ref{equation5}.  (b) Squares that are randomly oriented as
shown in Fig.~1, for which we estimate $\eta_c=0.9819\pm 0.0006$, so
$\phi_c=0.6254\pm 0.0002$.}
\label{Figure2}
\end{figure}

For square objects aligned parallel to each other in the 2-D system we
find 
\begin{mathletters}
\begin{equation}
\eta_c=1.098\pm 0.001
\label{equation6a}
\end{equation}
or, from (\ref{equation5}),
\begin{equation}
\phi_c=0.6666\pm 0.0004
\label{equation6b}
\end{equation}
\end{mathletters}
(see Fig.~\ref{Figure2}a). Our value of $\phi_c$ is within the error
bars of two previous determinations by Monte Carlo techniques, where
$\phi_c=0.668\pm 0.003$ \cite{King90} and $\phi_c=0.65\pm 0.02$
\cite{Garboczietal91}.  However our determination of $\phi_c$ is
slightly lower than that calulated by \cite{Alonetal90} whose Monte
Carlo simulations produced $\phi_c=0.6753\pm 0.0008$, and whose
application of the direct-connectedness expansion method yielded
$\phi_c=0.6912$.  In contrast, our value of $\phi_c$ is significantly
higher than the experimental one of \cite{DubsonGarland85} whose average
for 9 trials is $\phi_c=0.613\pm 0.013$.

We find that for randomly oriented square objects in 2-D
\begin{mathletters}
\begin{equation}
\eta_c=0.9819\pm 0.0006
\label{equation6c}
\end{equation}
or
\begin{equation}
\phi_c=0.6254\pm 0.0002
\label{equation6d}
\end{equation}
\end{mathletters}
(see Fig.~\ref{Figure2}b).  This is the first determination of these
values for randomly oriented squares.

These values for the continuum percolation threshold for aligned and
randomly oriented squares are lower than for discs, $\phi_c=0.67$
\cite{XiaThorpe88}, by a maximum of $\approx 0.5$\% for aligned squares
and $\approx 7$\% for randomly oriented squares.  We attribute the
significant difference in $\phi_c$ between discs and randomly oriented
squares to the possibility of randomly oriented squares intersecting
other squares whose centers are located at distances up to the diagonal
length of the square (see Fig.~\ref{Figure1}), whereas two discs can
only intersect if their centers are no further than one diameter away
from each other.  The similarity of $\phi_c$ for aligned squares and
discs may occur because both objects can only possibly intersect other
objects whose centers are separated at most by either the edge length of
the square or the diameter of the disc.
 
\subsection{Three Dimensions}

\begin{figure}
\centerline{
\epsfxsize=8.0cm
\epsfclipon
\epsfbox{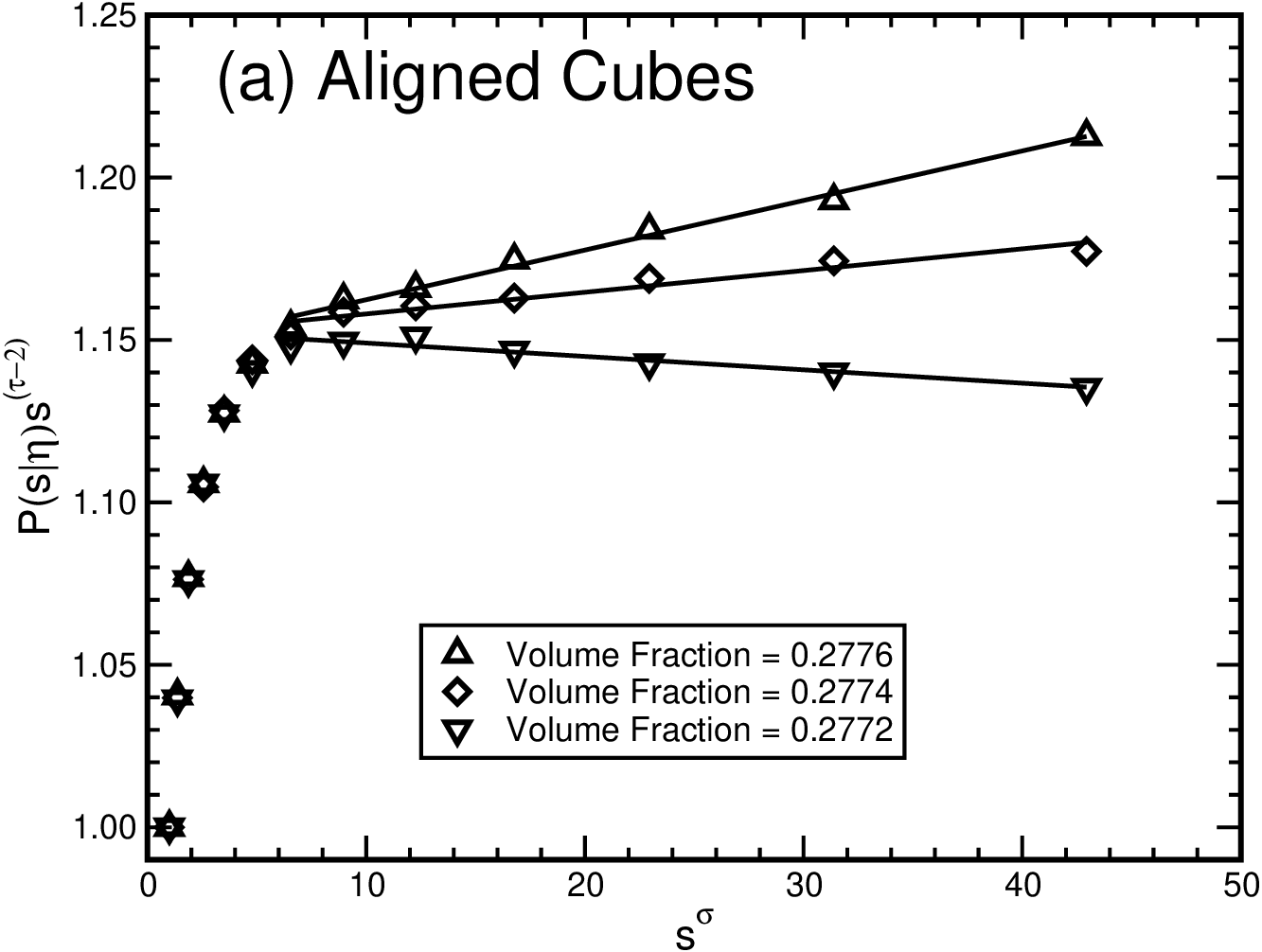}
}
\centerline{
\epsfxsize=8.0cm
\epsfclipon
\epsfbox{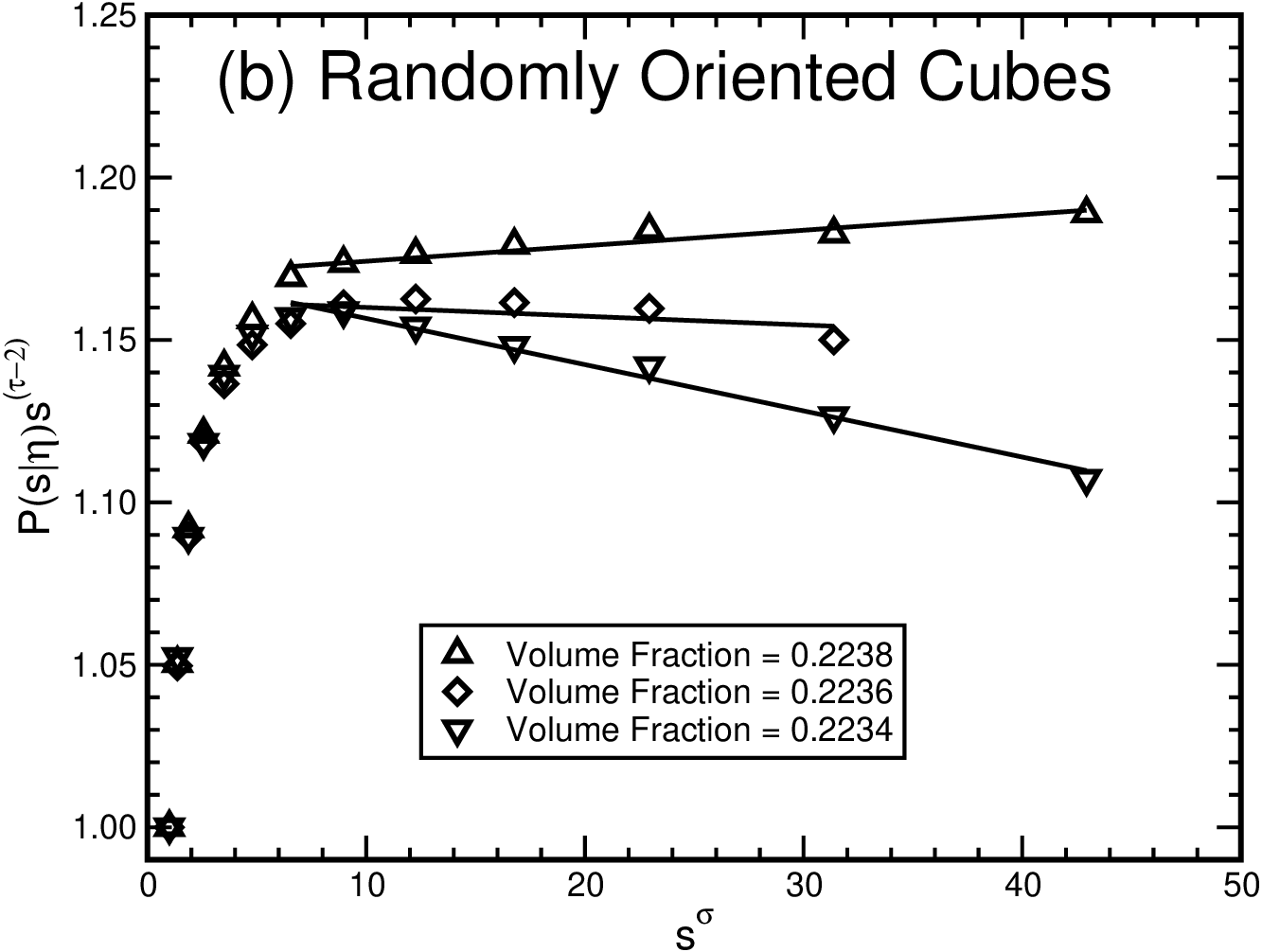}
}
\caption{Power-law scaled plots for determination of percolation
threshold for cubes of unit size in a $161\times 161\times 161$ 3-D
system based upon 50,000 realizations for each volume fraction.  At the
threshold $\eta=\eta_c$, $P(s|\eta)s^{\tau-2}$ is independent of
$s^\sigma$, which allows for accurate determination of $\eta_c$, which
is related to $\phi_c$ by Eq.~\protect\ref{equation5}. (a) Cubes whose
faces are aligned parallel to each other, for which case we estimate
$\eta_c=0.3248\pm 0.0003$, so by (\protect\ref{equation5})
$\phi_c=0.2773\pm 0.0002$.  (b) Cubes that are randomly oriented, for
which we estimate $\eta_c=0.2531\pm 0.0003$, so $\phi_c=0.2236\pm
0.0002$.}
\label{Figure3}
\end{figure}

For cubic objects aligned parallel to each other in the 3-D system,
\begin{mathletters}
\begin{equation}
\eta_c=0.3248\pm 0.0003
\label{equation7a}
\end{equation}
or
\begin{equation}
\phi_c=0.2773\pm 0.0002
\label{equation7b}
\end{equation}
\end{mathletters}
(see Fig.~\ref{Figure3}a).  The precision of this result is greater than
the most precise previous determination $\phi_c=0.280\pm 0.005$
\cite{Alonetal90}.  The critical volume fraction is significantly less
when cubic objects are allowed to have random orientations,
\begin{mathletters}
\begin{equation}
\eta_c=0.2531\pm 0.0003
\label{equation7c}
\end{equation}
or
\begin{equation}
\phi_c=0.2236\pm 0.0002
\label{equation7d}
\end{equation}
\end{mathletters}
(see Fig.~\ref{Figure3}b).  The result for randomly oriented cubes is
the same for cubes of edge-length $1/\sqrt 3$, 0.75, and 1.  Thus, as
expected, the percolation threshold is independent of the cube and
system size used.  This value is the first determination of $\phi_c$ for
the continuum percolation of randomly oriented cubes.

Comparision of the critical volume at the percolation threshold for
aligned cubes with that determined for spheres, $\phi_c=0.289573\pm
0.000002$ \cite{Lorenz01}, demonstrates that the difference in shape
between spheres and cubes affects $\phi_c$ by $\approx 4$\%
\cite{Alonetal90}.  Allowing cubes to randomly orient lowers $\phi_c$
$\approx 23$\%.  The difference between the randomly oriented cubes and
spheres is due to the same process as discussed above for discs and
squares, but in this case it is the greater length of the body diagonals
of cubes compared to the diameter of spheres or the edge length of
aligned cubes that enhances the probability of connectedness for
randomly oriented cubes at any given volume fraction.

\section{Discussion}

The continuum percolation threshold can be predicted with excluded
volume theory \cite{Alonetal90,Balbergetal84,Balberg87}: 
\begin{equation}
N_cV_{\mbox{\scriptsize ex}} = B_c,
\label{equation8}
\end{equation}
where $N_c$ is the critical density of objects \cite{Text2},
$V_{\mbox{\scriptsize ex}}$ is their excluded area or volume, and $B_c$
is the average number of bonds per object \cite{Balberg85}.  Originally,
$B_c$ was thought to be one constant for all parallel (i.e., not
randomly oriented) convex objects in 2-D and another constant in 3-D
\cite{Balberg85}, but later $B_c$ was determined to be different for
spheres and for cubes in 3-D \cite{Alonetal90}.  The excluded area for
discs and aligned squares of unit area is 4 and for randomly oriented
unit squares 4.084 \cite{Balbergetal84}.  For both spheres and aligned
cubes $V_{\mbox{\scriptsize ex}}$ is equal to 8 times their volume in
3-D.  For randomly oriented cubes $V_{\mbox{\scriptsize ex}}$ is 11
times their volume \cite{Kihara53}. Calculated values of $B_c$ in 2-D
and 3-D are presented in Table I.

In 2-D we determine $B_c= 4.39\pm 0.01$ for aligned squares and for
randomly oriented squares $B_c = 4.01\pm 0.01$.  The value for aligned
squares is similar to that originally proposed by Balberg for discs and
squares, $B_c = 4.5\pm 0.1$ \cite{Balberg85} and to the values for discs
and squares calculated from Monte Carlo simulations: $B_c= 4.43$
\cite{XiaThorpe88} and $B_c= 4.5\pm 0.1$ \cite{Alonetal90},
respectively.  On the other hand our value for $B_c$ is somewhat lower
than calculated by a series expansion technique, $B_c$ = 4.7
\cite{Alonetal90}.  Thus we confirm that $B_c$ has the same value,
within error, for discs and for aligned squares in 2-D.  However, a
different behavior is observed when squares are randomly oriented and
$B_c$ drops by $\approx 10 $\%, consistent with expectations that
non-parallel objects should exhibit a lower $B_c$ than parallel objects
\cite{Balbergetal84}.

Our Monte Carlo simulations yield $B_c = 2.59\pm 0.01$ for aligned
cubes, as is expected because of the agreement between our estimate of
the percolation threshold and previous estimates.  For randomly oriented
cubes $B_c=2.78\pm 0.01$, within error of $B_c$ for the continuum
percolation of spheres\cite{Alonetal90}.  This result is surprising in
light of the observation that $B_c$ for randomly oriented squares is
$\approx 10$\% below aligned squares and is contrary to expectations
that randomly oriented objects should have lower values of $B_c$ than
aligned ones \cite{Balbergetal84}. However, $B_c$ for randomly oriented
cubes does not exceed the limiting value predicted by the excluded
volume theory of the continuum percolation threshold \cite{Balberg85}.

Our results confirm previous research demonstrating the effect of object
shape on the threshold for continuum percolation.  We furthermore find
that the incorporation of random orientations of objects in continuum
percolation simulations significantly affects the percolation threshold.
Most of these effects are predicted by the application of excluded
volume theory to the calculation of the percolation threshold but,
surprisingly, allowing squares to randomly orient decreases $B_c$ in 2-D
whereas random orientation of cubes increases $B_c$ in 3-D.

\subsubsection*{Acknowledgements}

We thank P. Bourke, R. Consiglio, L.R. da Silva and S. Havlin for
discussions and BP, NSERC, and NSF for support.

\end{multicols}

\newpage

\begin{table}
\begin{tabular}{l|ccccc}
 \hline
Object & $V_{ex}$ for unit object & $\phi_c$ & $N_c$ & $B_c$
(calculated) & $B_c$ (literature)\\ \hline
Discs          & 4     & 0.67\protect\cite{XiaThorpe88}   & 1.1087 & 4.43  &
$4.5\pm 0.1$ \protect\cite {Balberg85}\\
 &&&&& $4.7$\protect\cite{Alonetal90}  \\
Aligned squares & 4     & $0.6666\pm 0.0004$ & $1.098\pm 0.001$ &
$4.39\pm 0.01$  & $4.5\pm 0.1$ \protect\cite {Balberg85} \\
 &&&&& $4.7$ \protect\cite {Alonetal90}     \\
Random squares  & 4.084 & $0.6254\pm 0.0002$ & $0.9819\pm 0.0006$ &
$4.01\pm 0.01$  &       \\ \hline 

Spheres        & 8  & 0.289573\protect\cite{Lorenz01} & 0.341889 & 2.74 & 2.79
\protect\cite{Alonetal90}\\ 
Aligned cubes & 8  & $0.2773\pm 0.0002$  & $0.3248\pm 0.0003$ &
$2.59\pm 0.01$  & 2.60 \protect\cite{Alonetal90} \\ 
Random cubes   & 11  & $0.2236\pm 0.0002$  & $0.2531\pm 0.0003$ &
$2.78\pm 0.01$  &  \\ 
\end{tabular}
\end{table}

\newpage

\newpage

\newpage

\newpage


\begin{references}
\bibitem{Text1} Three different sizes of objects were used in the 3-D
simulations to confirm the independence of the percolation threshold
from both object and system size. Unfortunately our range of system and
object size in both 2-D and 3-D simulations is constrained by computer
memory restrictions and CPU time available.
\bibitem{Text2} Here $N_c=\eta_c$ because the object volume equals 1.
\end{references}
\end{document}